# Flexible fiber batteries for applications in smart textiles

Hang Qu*[3], Jean-Pierre Bourgeois*[1], Julien Rolland[1], Alexandru Vlad[2], Jean-François Gohy[1] and Maksim Skorobogatiy[3]†

1. Institute of Condensed Matter and Nanosciences (IMCN), Université catholique de Louvain, Place L. Pasteur 1, B-1348 Louvain-la-Neuve, Belgium

2. Institute of Information and communication Technologies, Electronics and Applied Mathematics, Electrical Engineering, Université catholique de Louvain, Louvain la Neuve, B-1348 Belgium

3. Ecole Polytechnique de Montréal, Department of Engineering Physics, C.P. 6079, succ. Centre-ville, Montréal (Québec), Canada H3C 3A7

† Corresponding author
* These authors contributed equally to the paper

**ABSTRACT**

Here we discuss two alternative approaches for building flexible batteries for applications in smart textiles. The first approach uses well-studied inorganic electrochemistry (Al-NaOCl galvanic cell) and innovative packaging in order to produce batteries in a slender and flexible fiber form that can be further weaved directly into the textiles. During fabrication process the battery electrodes are co-drawn within a microstructured polymer fiber, which is later filled with liquid electrolyte. The second approach describes Li-ion chemistry within solid polymer electrolytes that are used to build a fully solid and soft rechargeable battery that can be furthermore stitched onto a textile, or integrated as stripes during weaving process.

**INTRODUCTION**

With the rapid development of micro and nanotechnologies and driven by the need to increase the value of conventional textile products, fundamental and applied research into smart textiles has recently flourished [1-3]. Generally speaking, textiles are defined as "smart" if they can respond to the environmental stimulus, such as mechanical, thermal, chemical, electrical, and magnetic. Many applications of "smart" textiles stem from the combination of textiles and electronics (e-textiles). Most of the "smart" functionalities in the early prototypes of e-textiles were enabled by integrating conventional rigid electronic devices into a textile matrix. The fundamental incompatibility of the rigid electronic components and a soft textile matrix create a significant barrier for spreading of this technology into wearables. This problem motivated many recent efforts into the development of soft electronics for truly wearable smart textile. This implies that the electronic device must be energy efficient to limit the size of the battery used to power it. Needless to say that to drive all the electronics in a smart textile one needs an efficient, lightweight and flexible battery source. Ideally, such a source will be directly in the form of a fiber that can be naturally integrated into smart textile during weaving.

In this paper we report two alternative approaches towards producing soft batteries for smart textile application. One approach uses relatively simple inorganic electrochemistry and innovative packaging, while another one uses a more complex Li-ion chemistry within solid polymer electrolytes and relatively simple battery geometry.

**1. BATTERIES BASED ON Al-NaOCl CHEMISTRY AND INNOVATIVE PACKAGING**

Traditional galvanic battery cells, which were discovered centuries ago, still show their robust vitality in the scientific and industrial world due to their ease of operation and simplicity of electro-chemistry. Among a large number of galvanic cells, the Al-NaOCl system has been comprehensively studied due to its advantages such as cost-effectiveness, availability of the chemical components and high current-carrying capacity [4, 5].

In this paper we first show how a generally bulky galvanic cell can be made in a slender fiber form by using fiber drawing technique. The fiber battery is fabricated from rods of low density polyethylene (LDPE) that features good chemical resistance to aqueous solutions of salt, acids and alkalis, and a large number of organic solvents. To fabricate the fiber, we first drill three intercommunicating holes in the LDPE rod to form a fiber perform. Then, the preform is co-drawn in the drawing tower with aluminum (99% pure) and copper wires (99% pure). Up to several km of battery fiber can be produced in a single run using our research-grade drawing tower. The cross section of a typical ~1mm diameter fiber battery (without electrolyte) is shown in Fig. 1.

The central hole that separates two metallic electrodes is then filled with NaOCl (bleach) electrolyte. The capacity of such fiber-based batteries is measured to be ~10 mAh per 1 meter of fiber. We note that a key factor that limits performance of our fiber battery is generation of the hydrogen gas, which eventually displaces the electrolyte from the central hole, thus interrupting the normal functioning of the battery. The advantages of the proposed fiber battery are low cost and commercial availability of all the components, capability of mass production, the lack of poisonous materials, and simplicity in electro-chemistry. More importantly, due to its flexibility and light weight, the fiber battery can be weaved into a wearable battery textile or integrated into other "smart" textiles as power source, which allows the fiber battery to find its niche market in many scientific and industrial applications.

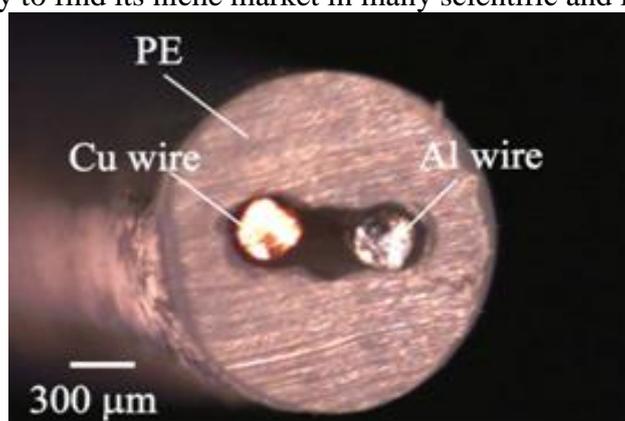

Fig. 1. Cross section of a fiber battery (before filling it with electrolyte). An aluminum wire and a copper wire are inserted into the two extreme channels of the polyethylene fiber, while the third hole separating the fibers is later will with electrolyte.

**1.1 Theory**

For the Al-NaOCl fiber battery system, the redox potentials, $E^0$, versus Standard Hydrogen Electrode (SHE) are described as follows [4]:

$$\text{Anode: } Al \rightarrow Al^{3+} + 3e^-; \quad E^0 = 1.7V, \quad (1)$$

$$\text{Cathode: } OCl^- + H_2O + 2e^- \rightarrow 2OH^- + Cl^-; \quad E^0 = 0.81V, \quad (2)$$

Thus, the most likely overall reaction of the fiber battery is:

$$\text{Overall reaction: } 2Al + 3NaOCl + 3H_2O \leftrightarrow 3Na^+ + 3Cl^- + 2Al(OH)_3 \downarrow, \quad (3)$$

Reactions (1)-(3) demonstrates that the Al-NaOCl system is capable of attaining a theoretical potential of 2.51 V at "standard chemical state". At the same time, the following parasitic reaction leads to the release of the hydrogen gas during battery operation [5].

$$\text{Parasitic reaction: } 2Al + 2H_2O + 2OH^- \leftrightarrow 2AlO_2^- + 3H_2 \uparrow, \quad (4)$$

The hydrogen gas generated in the parasitic reaction would expel the electrolyte out of the fiber battery, thus degrading the performance of the fiber battery. Besides, we note that the product of reaction (3) is $Al(OH)_3$, which is a gel-like precipitant [6] that may also lead to the degradation of the fiber performance by blocking the ion flow between the two electrodes. The copper wire in this system does not participate the battery reactions, and only acts as a current collector.

**1.2 Experimental details, galvanic cell**

Before characterizing fiber batteries, we first study performance of a standard Al-NaOCl galvanic cell. The galvanic cell is assembles by attaching two 7cm-long aluminum and copper wires to glass plates, as shown in Fig.2. The two electrodes are then placed in a reservoir filled with solution of NaOCl (electrolyte). Two different concentration of NaOCl were used in our experiments, which were 5% and 14.5%. At each concentration of electrolyte, the open-circuit voltage $V_o$, and a short-circuit current $i_m$ are measured as a function of separation between the two electrodes (0.5 cm, 1 cm, 2 cm, 3 cm and 4 cm). This enables us to estimate the internal resistance of the galvanic cell (by dividing $V_o$ by $i_m$) as a function of the distance between electrodes.

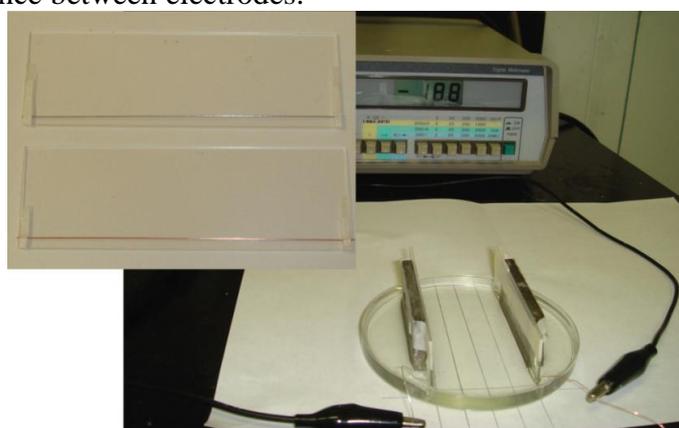

Fig. 2. Setup for measuring the short-circuit currents and the open voltages of an Al-NaOCl galvanic cell. The inset shows two glass plates to which the copper and the aluminum wires are attached.

The open-circuit voltage of the galvanic cell is measured to be 1.64 V using the electrolyte of 14.5% NaOCl solution and 1.44 V using the electrolyte of 5% NaOCl solution. Moreover, the open-circuit voltage is largely insensitive to distances between the two electrodes. In Fig. 3, we show time evolution of the short-circuit current of the galvanic cell as a function of the inter-electrode distance 0.5 cm, 1 cm, 2 cm, 3 cm and 4 cm.

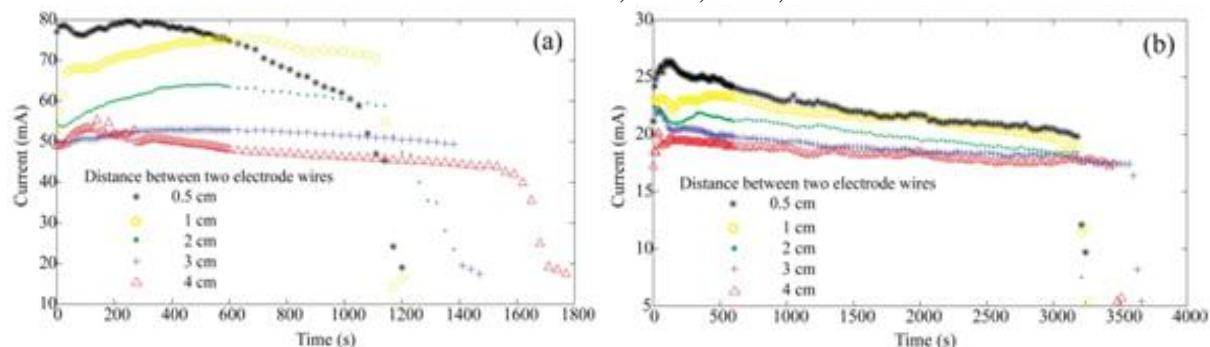

Fig.3. Short-circuit current in the Al-NaOCl galvanic cells. The distance between the two electrode wires is set to be 0.5 cm, 1 cm, 2 cm, 3 cm, and 4 cm. The electrolytes used in the battery cells are 5 wt% NaOCl solution in (a) and 14.5 *w*t% NaOCl solution in (b).

First, we observe that short-circuit current increases significantly with the concentration of NaOCl electrolyte. Secondly, from Fig. 3 we see that the short-circuit current generally increases in the first several minutes after the battery discharging, which is likely due to fact that the oxidation layer exiting on the Al surface is gradually dissolved by the electrolyte solution. Then, the current gradually decrease because of the production of the hydrogen gas and the gel-like $Al(OH)_3$ (see Eq. (3) and Eq. (4)) that concentrate in the vicinity of the electrodes and reduce their effective area. Finally, the failure mode of the battery is usually the brakeage of the Al wire.

The data in Fig. 3 enables us to calculate the internal resistance of the galvanic cell by dividing $V_o$ by $i_m$, which is presented in Fig. 4. We estimate $i_m$ by averaging the current within the first 10 minutes after the battery discharging. Internal resistance $R$ of an infinite electrolyte featuring two cylindrical electrodes separated by distance $d$ can be easily found from basic electrostatics [7]:

$$R\delta = (1/\pi L)\ln(d/r_0 - 1), \quad (5)$$

where $\delta$ is the conductivity of electrolyte, $L$ is the length of the electrodes, $r_0$ is the diameter of the wires. Comparison of the measured resistance and logarithmic fit is presented in Fig. 4. We note, however, a notable deviation from the theoretical prediction and experimental observation. The reason for this difference is still not quite clear to us.

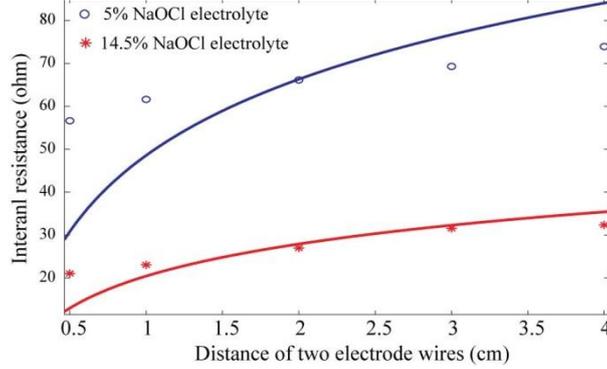

Fig. 4. Internal resistance of galvanic cell calculated by dividing $V_o$ by the short-circuit current $i_m$. In stars and circles we present experimental date, while in solid line we present a logarithmic fit (5).

### 1.3 Experimental details, battery fiber

In the second experiment, we characterize performance of the two 10cm-long fiber batteries. We chose two batteries with the inter-electrode distances of 400 and 650 microns. The open-circuit voltage of both batteries was measured to be ~1.5V. Then, we studied battery generated current under different resistive loading conditions, such as: short-circuited battery, resistive load of 33-ohm and resistive load of 333-ohm. The time evolution of current under different loading conditions is shown in Fig. 5.

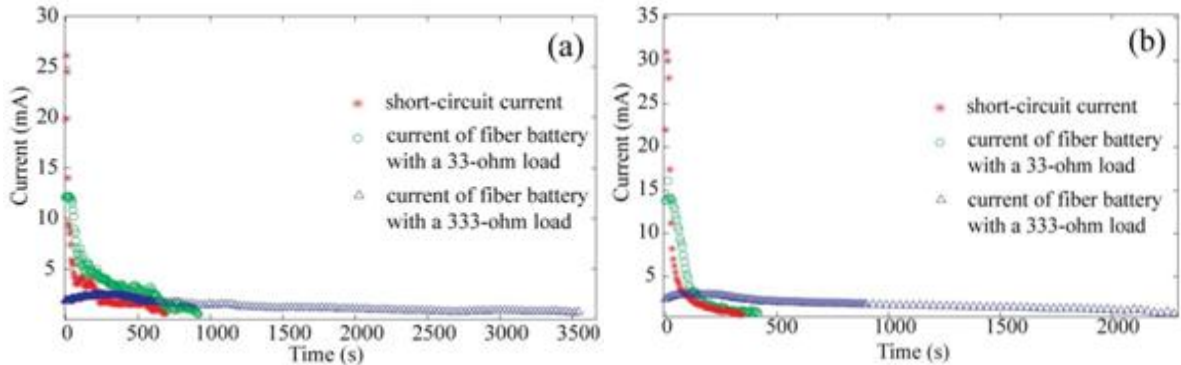

Fig. 5. Current generated by the fiber batteries under different loading conditions. The inter-electrode distance in the battery is ~ 400 micron (a) and ~ 650micron (b).

From Fig. 5 we conclude that the normalized internal resistance of our fibers is ~ 5Ω·m. Moreover, from Fig. 5 we can calculate the fiber battery capacity by integrating the current. Thus, the normalized capacity of these two fiber samples per 1 meter of battery length is ~$10^{-2}$ Ah/m and is largely independent of the loading conditions.

## 1.4 Discussion

As we have mentioned in the introduction, development of the fiber batteries is motivated by their applications in smart textiles. Due to their small <1 mm diameter, and mechanical robustness, such fibers can be densely weaved into a traditional textile. For example, it is quite easy to weave 200 pieces of 10-cm long fibers into a 10cm×50cm ribbon. By connecting every 2 adjacent batteries in series, and then connecting the resulting 100 fiber pairs in sequence would result in a 3V battery with a total capacity of 0.1Ah. This textile-based battery can light up a standard light emitting diode (25mA@3V) for about 5h.

Finally, we note that one of the main limitations in the operation of our fiber battery is the generation of hydrogen gas. As shown in Fig. 6, gas bubbles are generated immediately after the current starts flowing through the battery (Fig.6 (b)), and a large portion of the electrolyte solution is expelled out the fiber within the first minute after discharging (Fig.6 (c)). In principle, this drawback can be alleviated by adding certain chemicals to absorb the hydrogen gas or convert hydrogen into water through hydrogen-oxygen reaction.

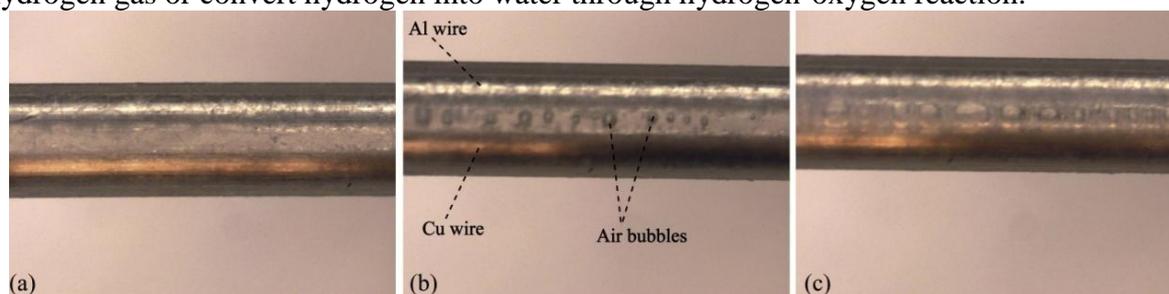

Fig.6. Fiber battery (a) open circuit; (b) short-circuited, after 5s; (c) short-circuited, after 40s.

## 2. BATTERIES BASED ON LI-ION CHEMISTRY AND SOLID IONIC ELECTROLYTES

Due to their design and composition, lithium ion batteries allow the production of reliable power sources featuring low cost and high energy capacity [8-10]. In order for Li-ion batteries to be better suited for smart textiles, ideally, such batteries should be flexible and they should not contain any liquid electrolyte. In our recent work [11] we have reported fabrication of flexible and stretchable batteries composed of strain free $LiFePO_4$ cathode, $Li_4Ti_5O_{12}$ anode and a solid poly ethylene oxide (PEO) electrolyte as a separator layer. Featuring solid thermoplastic electrolyte as a key enabling element this battery is potentially extrudable or drawable into fibers or thin stripes which are directly compatible with the weaving process used in smart textile fabrication. The paper first detailed the choice of materials, fabrication and characterization of electrodes and a separator layer. Then the battery was assembled and characterized, and finally, a large battery sample made of several long strips was woven into a textile, connectorized with conductive threads, and characterized. Two practical aspects of battery design were investigated in details: first was making composites of cathode/anode material with optimized ratio of conducting carbon and polymer binder material, and second was characterization of the battery performance including cycling, reversibility, and compatibility of the cathode/anode materials.

In Fig. 7 we present an example of a textile battery comprising 8 flexible battery stripes woven together and connectorized in series to power up a 3 V light-emitting diode (LED). This battery provides dim LED light for several hours and it can be recharged. The electrode compositions used in this sample contained high content of PEO (at least 50%).

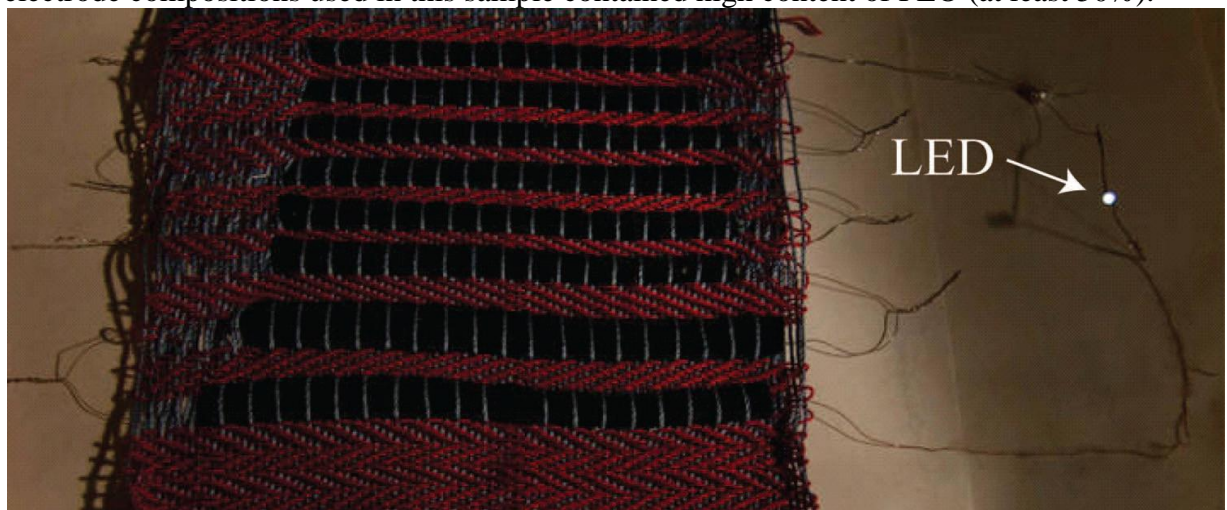

Fig.7. Textile battery is made of 8 Li-ion flexible battery stripes woven with cotton thread and connectorized in series using copper and aluminum wires (one per stripe per side) as electron collectors. The resultant battery is powerful enough to light up a 3 V LED for several hours (see [8]).

In our original work [11] on flexible solid Li-ion batteries we have mostly focused on using aqueous solutions of PEO rather than solutions of PEO in organic solvents in order to fabricate both the electrode binders and separator layers. In fact, we were hoping to develop an environmentally friendly process for the electrode and polymer electrolyte fabrication. As a result, battery performance has significantly suffered, and open circuit voltage was much smaller than the theoretically expected. Consequently, we are in the process of changing the processing conditions for the battery assembly in order to improve its performance.

Particularly, we are now focusing on using organic solvents for our cathode/anode and separator layer fabrication. In what follows we are describing our first measurements of the ionic conductivity of solid PEO and Li-salt mixtures as a function of the salt concentration. The salt-impregnated PEO plays the role of a solid ionic electrolyte, which is a critical element in the battery design. In fact, for applications in soft textile batteries we expect that solid electrolytes should not only have as high ionic conductivity as possible, but also maintain the desired mechanical properties such as flexibility, softness, etc.

**2.1 Experiment details**

Poly (ethylene oxide) (POE) of molecular weight of 400.000 g/mol was dissolved in acetonitrile at 50°C for 2 days. Appropriate amount of lithium salts $LiClO_4$ were added and the solutions were stirred for at least 3 hours. Solutions were then poured in aluminum dishes and evaporated under reduced ambient pressure to cast PEO films. The films were finally dried for 48hours at 30°C under vacuum. Chemicals were obtained from Alfa Aesar and used without further purification.

Ionic resistance of PEO films was obtained via electrochemical impedance spectroscopy measurements that were carried out using an Autolab spectrometer. The test cell comprised two copper or aluminum electrodes with area of ~ 7,07cm². PEO film surface imaging and cross section thickness measurements were performed using optical microscope.

Finally, ionic conductivity of PEO and Li-salt mixtures were calculated using Ohm's law for the resistance of a thin film:

$$\sigma = \frac{L}{R\,S}, \quad (6)$$

where *R* is the ionic resistivity obtained via impedance spectroscopy measurements, *L* is the film thickness (as measured by the optical microscope) and *S* is the film area.

## 2.2 Mechanical properties of films based on PEO / Li-salt mixtures

PEO films can change significantly their mechanical properties as a function of the lithium salt concentration (additive to PEO). By adding increasing amounts of salt to PEO films, we intend to decrease the PEO crystallization which, in turn, increases their ionic conductivity. The difference between crystalline PEO film and amorphous PEO film is clearly visible in Fig. 1. In fact, PEO films with lower concentrations of salt are opaque (high degree of crystallinity), while at the same time they are relatively rigid and plastic-like. PEO films with higher concentration of salt result in jelly-like transparent films (Fig. 1(b)), which are difficult to detach from the aluminum dishes without ripping.

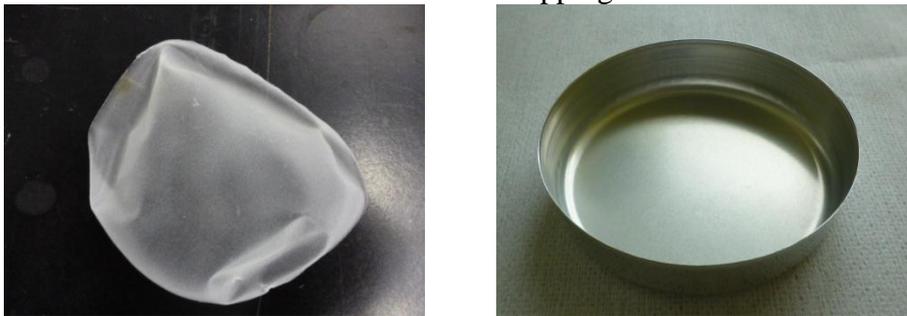

Fig. 8. PEO films with different concentration of Li-salt: a) 1:20 $Li^+$/PEO ratio results in highly crystalline opaque films, b) 1:5 $Li^+$/PEO results in jelly-like transparent films.

## 2.3 Electrochemical impedance spectroscopy

The first test was to make sure that the salt-impregnated PEO films were electrically insulating, which was verified directly using a multi-meter. Furthermore, to find ionic conductivity of the PEO film electrochemical impedance spectroscopy (EIS) was then performed in the 1Hz-1MHz range under a 20mV potential. During measurements the PEO films were sandwiched between two cylindrical copper electrodes. Results of the EIS measurements are typically presented in the form of a Nyquist plot (see Fig. 9). In such a plot imaginary part (Z'') of the complex impedance of an electric circuit under study is plotted as a function of its real part (Z').

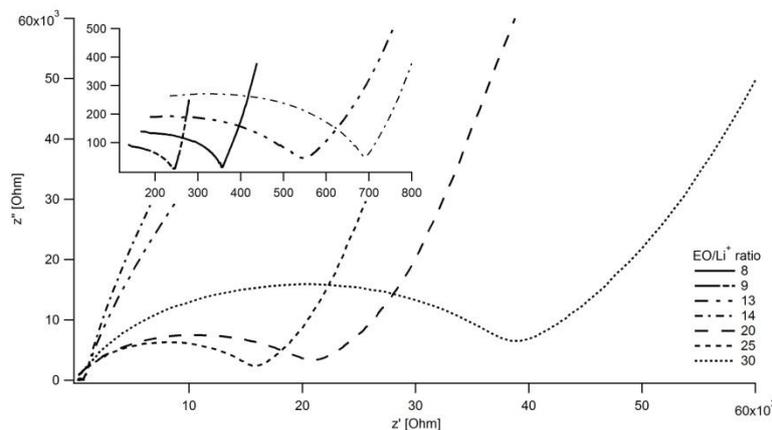

Fig. 9. Nyquist plot of EIS performed on PEO film with different Li-salt concentrations

## 2.4 Data interpretation

In order to interpret the results of the impedance spectroscopy (Fig. 9) we have to assume a certain model for the equivalent electrical circuit that describes our measurement cell (PEO film sandwiched between two cylindrical metal electrodes). One of the possible models used in our studies is presented in Fig. 10.

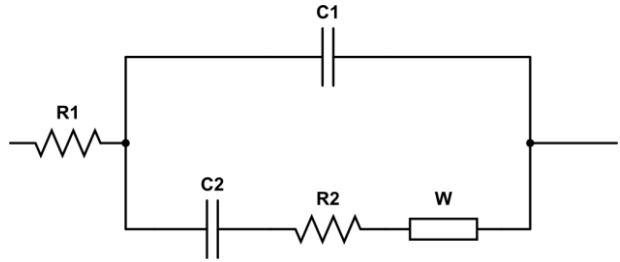

Fig. 10: Electrical equivalent circuit chosen

| Symbol | Interpretation | Comments |
| --- | --- | --- |
| C1 | Geometrical capacitance | Capacitance resulting from the two copper electrodes sandwiching the insulating PEO film |
| R1 | Circuit resistance | Representing the copper electrodes and the connection wires |
| C2 | Electrode/electrolyte interfacial capacitance | Capacitance resulting from the contact between copper electrode and electrolyte. |
| R2 | Polymer electrolyte resistance | <u>Ionic resistance of the electrolyte (the quantity of main interest to us)</u> |
| W | Warburg impedance | Representing ionic diffusion in the electrolyte $W=\frac{1}{Y_0\sqrt{j\omega}}$, $\omega$ is the angular frequency, $Y_0$ is a Warburg constant |

The analytical solution for the complex impedance of the electrical circuit presented in Fig. 10 is given by:

$$Z = \frac{1+jR_1\omega C_1+\frac{R_1}{R_2+\frac{1}{Y_0\sqrt{j\omega}}+\frac{1}{j\omega C_2}}}{j\omega C_1+\frac{1}{R_2+\frac{1}{Y_0\sqrt{j\omega}}+\frac{1}{j\omega C_2}}}. \quad (7)$$

This analytical solution is then used to fit the Nyquist plot recorded using EIS and to find ionic resistance of the film defined by the R2 parameter.

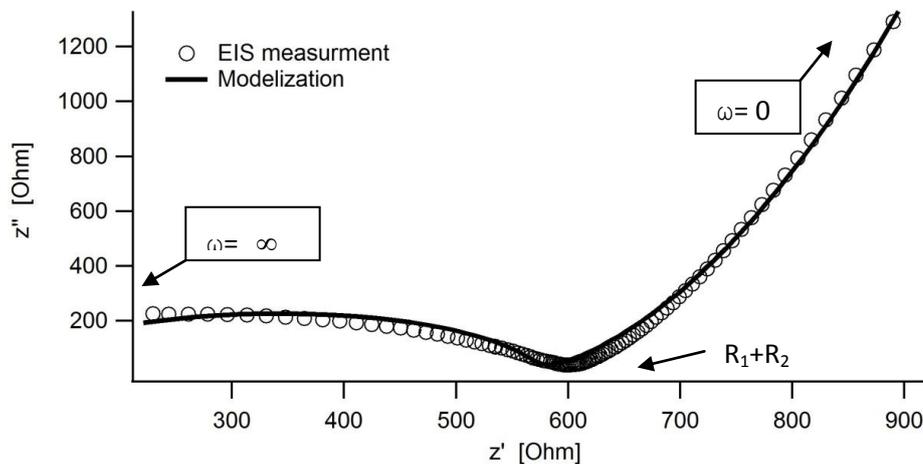

Fig. 11. A typical experimental Nyquist plot (open circles), Nyquist plot for the equivalent circuit (solid line)

## 2.5 Results and discussion

In Fig. 12 we present the main result of our work, which is dependence of the ionic conductivity of PEO / Li-salt mixtures on the Li ion concentration. We also find that solid films usually have lower ionic conductivities (~$10^{-7}$-$10^{-6}$ S/cm) compared to the jelly-like films (~$10^{-5}$ S/cm), which is a simple manifestation of the effect of salt concentration on the film mechanical properties. These values are also consistent with those published in [12].

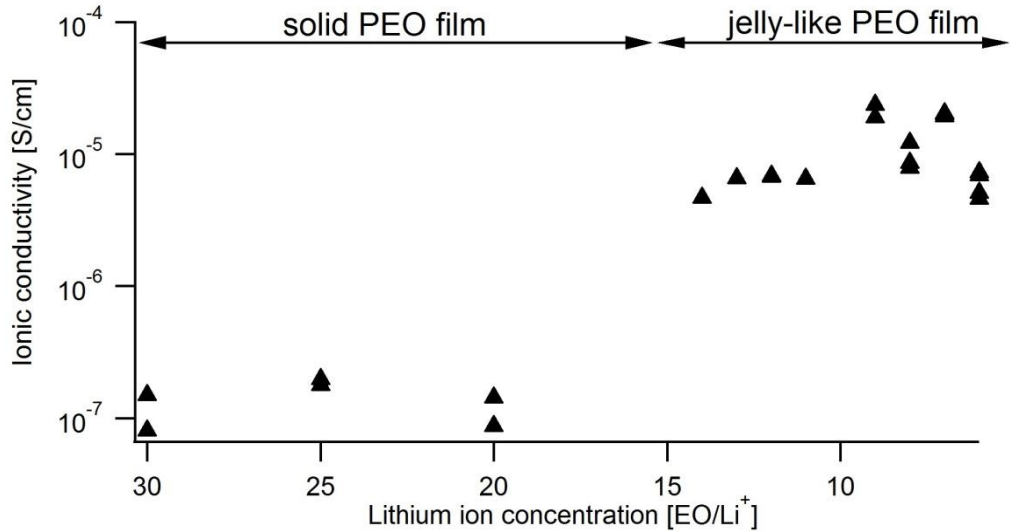

Fig. 12. Ionic conductivity of PEO films versus lithium ion concentration

## CONCLUSION

In summary, we propose and demonstrate two soft batteries for smart textile applications. The first one is a fiber battery that features Al and Cu electrodes in a polyethylene microstructured fiber filled with NaOCl electrolyte. The performance of the fiber battery is studied and compared to a bulk galvanic cell of the same type. The capacity of a 1-meter long fiber battery is measured to be ~$10^{-2}$Ah. The fiber batteries are light-weight, low cost, and environmentally safe.

The second flexible battery for smart textile applications has been demonstrated by us in [11] where we have used conventional Li battery materials including LiFePO4 cathode, $Li_4Ti_5O_{12}$ anode and PEO solid electrolyte. By introducing large quantities of the thermoplastic PEO binder in the battery electrodes and separator layer one can potentially realize a fully extrudable/drawable battery system, which could allow direct drawing of battery fibers, which are ideal for textile applications. In this paper we have also presented our latest measurements of the ionic conductivity of the PEO / $LiClO_4$ mixtures as a function of the salt concentration. These solid electrolytes are in the heart of the flexible batteries, and their electrochemical and mechanical properties are strongly interrelated.

## ACKNOWLEDGMENTS

The authors would like to acknowledge Canada Research Chair of Prof. M. Skorobogatiy and the FNRS-FRIA contract No. FC93708 for providing financial support for this collaboration. We would like to thank Profs. L. Martinu and J. Sapieha for letting us using their impedance spectroscopy equipment.

The authors would like to acknowledge that development of the inorganic Al-NaOCl fiber batteries were mostly done by Q. Hung who had most fruitful discussions with J.P. Bourgeois on the subject, while the work on Li-ion batteries based on organic solvent processing was mostly done by J.P. Bourgeois within a Canada – Belgium collaboration.